\documentclass[12pt,letterpaper,hyper]{JHEP3}

\usepackage{epsfig}
\usepackage[utf8]{inputenc}
\usepackage{epic,eepic}
\usepackage{amsmath,epsfig}
\usepackage{amssymb,amsfonts}
\usepackage{graphicx}

\title{On Vector Meson Masses in a Holographic SQCD}

\preprint{DFTT/17/2010}

\author{
Aldo L. Cotrone$^{a,b}$, \,
Anatoly Dymarsky$^{c}$, \,
Stanislav Kuperstein$^{d,}$\footnote{cotrone@to.infn.it, dymarsky@ias.edu, skuperst@lpthe.jussieu.fr}\\

\it{$^a$ Institute for theoretical physics, K.U. Leuven; Celestijnenlaan 200D, B-3001 Leuven, Belgium.}\\

\it{$^b$ Dipartimento di Fisica Teorica, Universit\`a di Torino and I.N.F.N. - sezione di Torino
Via P. Giuria 1, I-10125 Torino, Italy.
}\\

\it{$^c$ School of Natural Sciences, Institute for Advanced Study, Princeton, NJ 08540.}\\

\it{$^d$ LPTHE Universit\'e Pierre et Marie Curie -- Paris 6; CNRS UMR
7589, Tour 13-14, 4 Place Jussieu, 75252 Paris Cedex 05, France.}}

\abstract{
We study probe flavor branes supersymmetrically embedded in the Klebanov-Strassler
background. The embedding is controlled by a single complex parameter
$\mu$ related to the quark mass. We study the spectrum of the vector mesons as a function of $\mu$
and compare it with the experimental data. Assuming that the $\rho(2150)$ meson is a $3\, ^3 S_1$ mode we
find a remarkable agreement with the masses of the lightest excited $\rho$-mesons.
Also, for a certain range of parameters our model exhibits an unusual behavior.  The mesons
built of the lighter quarks become more heavy than the ones built of the heavier quarks.
We comment on a possible connection this phenomenon may have with the masses of the putative pure
strange-anti strange mesons.
}


\renewcommand{\Im}{{\rm Im}}
\renewcommand{\Re}{{\rm Re}}

\def\bea{\begin{eqnarray}}
\def\eea{\end{eqnarray}}
\def\be{\begin{equation}}
\def\ee{\end{equation}}
\def\ba{\begin{align}}
\def\ea{\end{align}}
\def\bse{\begin{subequations}}
\def\ese{\end{subequations}}

\newcommand{\bem}{\begin{pmatrix}}
\newcommand{\eem}{\end{pmatrix}}

\def\sl{$SL(2,R)$}

\def\={\;  = \;}
\def\+{\, + \,}

\def\bar{\overline}

\def\rt2{\sqrt{2}}

\renewcommand{\Im}{\mbox{Im}}
\renewcommand{\Re}{\mbox{Re}}

\begin{document}

%\begin{displaymath}
%\xymatrix{
%\bullet \ar@/^/[r]
%\ar@/_/@{.>}[r] &
%\bullet
%}
%\end{displaymath}

\section{Introduction and Summary}

The gauge/string correspondence and its numerous generalizations (see \cite{Bigazzi:2003ui} for a review)
conjectures the equivalence between (super)string theory on a ten-dimensional space-time and
some gauge theory living on the boundary of
this space-time. The approach is based on taking the near horizon limit of the
background produced by a large number of $Dp$-branes. On the field theory side the number of branes
$N_c$ corresponds to the rank of the gauge group(s). Therefore the gauge/string correspondence provides
a holographic description of the field theory
in the planar limit. Large $N_c$ and supersymmetry
 contribute to the differences between the holographic models and the real world QCD.
Yet, for many phenomenological applications the holographic approach can provide a valuable insight.
In the holographic approach, in order to add a meson sector one has to include ``flavor" D-branes \cite{Karch:2002sh},
giving rise to the quark degrees of freedom.
Such models, colloquially called  holographic QCD, have been extensively studied over the last decade
for various gravity backgrounds with different flavor D-branes.

We choose the Klebanov-Strassler (KS) \cite{Klebanov:2000hb} background as the starting point for our
holographic setup. This background is dual to a confining supersymmetric gauge theory which cascades to a
$\mathcal{N}=1$ $SU(M)$ SYM in the deep IR.
On the gravity side the KS solution is based on the $6d$ deformed conifold. The deformation parameter
$\epsilon$ is related to the dynamical scale $\Lambda$ of the gauge theory.
Apart from $\Lambda$,  the real life QCD also includes additional massive parameters --
the quark masses $m_{\rm q}$.
In this paper we will focus on a holomorphic $D7$-brane previously studied in \cite{Kuperstein:2004hy}.
For the sake of simplicity, we will consider a single $D7$-brane.
Thus we can use the quenched approximation ignoring the backreaction
of the flavor brane on the gravity background. In general, the approximation is valid for
$N_{\rm f} \ll N_{\rm c} $.
The holomorphicity of the embedding implies that it breaks none of the background supersymmetries and
the dual gauge theory with flavor is a $\mathcal{N}=1$ SYM with fundamental matter. The embedding of the
$D7$-brane is parametrized by a single complex parameter $\mu$ which is related to the
quark mass. The preserved supersymmetry allows us to write down the precise form of the
superpotential. It appears that apart from the quark mass term there is a
term describing quartic interaction between the quarks and the KS bi-fundamentals.

The low energy fields living on the $D7$-brane are the complex scalar that corresponds to the directions
transversal to the brane and the $8d$ gauge field. From the $4d$ point of view the latter decomposes into
a gauge field and another three real scalars. The gauge field is dual to the conserved $U(1)_{\rm f}$ flavor
symmetry current that gives rise to the vector mesons. All other fields are dual to scalar
operators related to each other by supersymmetry. We review the meson multiplet structure  later in the text.
Although the calculation of the scalar meson spectrum currently appears to be a hard technical problem, the vector
meson spectrum can be computed relatively easy for any value of $\mu/\epsilon$.
We thus have a remarkable opportunity to investigate how the vector meson mass $m$
depends on both the quark mass $m_{\rm q}$ and the IR scale $\Lambda$,
with the former fixed by $\mu$ and the latter by $\epsilon$.

It is worth recalling here that we are dealing with a supersymmetric version of QCD  and as a result both
$\Lambda$ and $m_{\rm q}$ ($\epsilon$ and $\mu$) are complex parameters. Naively, for a fixed $\epsilon$,
the meson masses are expected to grow as we increase $\vert \mu \vert$: the meson built of heavier quarks is
supposed to be heavier.
Surprisingly, we found numerically that this is not the case if the ratio $x \equiv \mu/\epsilon$ is
real and smaller than a certain value, approximately equal to one.
In this case only the lowest meson behaves as expected, \emph{i.e.} its mass increases with   $x$. At the same time
the masses of all other modes decrease as $x$ increases from $0$ to $1$. Then, around $x \approx 1$ the masses reach
their minimum and increase afterwards.  This is an unusual behavior both because the meson mass is not a monotonic
function of $\mu$ and because it singles out the lowest mode. To confirm the result of numerics we
use the WKB analysis reaching the same conclusion.

In Nature the lightest vector mesons are the $\rho$-mesons built from the $u$ and $d$ quarks.
These quarks are much lighter than the QCD scale. We compared the ratio of the
lowest $\rho$-meson masses with the predictions of our model for $\mu=0$ and found a remarkably
good agreement assuming that the $\rho(2150)$ meson is a  $3\, ^3 S_1$ mode.\footnote{
The experimental status of the $3\, ^3 S_1$ mode is presently unclear. We discuss this issue later in the text.}

Next in the spectrum of vector mesons after the $\rho$'s  are the $\phi$-mesons built
from the heavier $s$ and $\bar{s}$ quarks. These mesons, however, can mix the ``pure" $s \bar{s}$
states with $u \bar{u}$ and  $d \bar{d}$. To  compare with the holographic model we use putative pure
``$\phi$"-mesons, whose masses can be derived from the spectrum of
the $K^{\star}$ and $\rho$-mesons. We found similarity between the non-monotonic behavior of the spectrum
as a function of $x$ in our model and the putative ``$\phi$"-meson masses.
Namely the first excited mode of the ``$\phi$"-meson (that consists of the heavy $s$-quarks) is lighter than
the first excited mode of the $\rho$-meson (that consists of the light $u,d$-quarks). Remarkably, this applies
only for the excited but not to the lowest mode, similarly to what we observed in our holographic model.

These optimistic results should not be taken for granted
since our model is only a distinct relative of the real-world QCD.
Nevertheless, we believe that the observed
similarities may not be a mere coincidence. It can rather suggest that the ratio of masses for certain
light excitations in the QCD-like theories is quite robust and is not sensitive neither to the number
of colors, nor to the explicit field content of the field theory. A similar phenomenon was previously
observed for certain glueballs: the KS theory happens to  predict with a reasonable accuracy the mass
ratio for the lightest scalar glueballs of the pure $SU(3)$ theory \cite{Dymarsky:2008wd}.

As another option, it may be the case
that the inverse mass phenomenon is actually a result of some  peculiarities of our model, not shared by QCD.
Thus the quartic interaction between the quarks and the original fields of the KS model may provide
a nice intuitive explanation for the different spectrum behavior for real and pure imaginary $\mu$.
This argument has also a clear geometrical counterpart. In terms of the dual geometry,
the minimal distance between the tip of the deformed
conifold and the lowest point on the ${\rm D}7$-brane depends on $x$ in an interesting way.
When $x$ has a non-zero phase the distance vanishes only for $x=0$, while for real $x$ the brane extends
all the way to the tip as long as $0\leqslant x \leqslant 1$.  Hence we have different behavior for real and complex $x$.

\subsection*{Organization of the paper}

In the next section we briefly review the Klebanov-Strassler (KS) model with the embedded  $D7$-brane. We also remind the reader the basics of the dual
gauge theory. Next we discuss the $D7$-brane world-volume fields and the meson multiplet structure.
In Section \ref{Numerics} we perform the calculation of the spectrum using the holographic approach.
We compute the spectrum using the ``shooting'' technique  and compare it with the results of the
WKB approximation. Then we provide an interpretation of the spectrum behavior based
on the peculiarities of our model. Section \ref{Experiment} is devoted to the comparison with the experimental data.

\section{The conifold based models: flavored and un-flavored} \label{review}

The deformed conifold is a regular six dimensional non-compact CY manifold defined by the equation
\be
\sum_{i=1}^{4} z_i^2  = \epsilon^2 \, .
\ee
%The singular conifold with $\epsilon=0$ is invariant under complex rescaling of the $z_i$'s and
%has $SU(2)\times SU(2)\times U_R(1)$ isometry.
%The deformation parameter $\epsilon$ breaks  the $U_R(1)$ symmetry and the scale invariance,
%producing a finite-size $S^3$ at the tip of the conifold.
%
The relevant $10d$ supergravity solution is of the GKP type \cite{Giddings:2001yu}
with constant dilaton and a warped metric
\be  \label{10dmetric}
d s^2_{10} = h^{-1/2} (r) d x_{\mu} d x^{\mu} + h^{1/2} (r) d s_6^2 \, .
\ee
Here $d s_6^2$ is the (deformed) conifold metric and the warp-factor $h(r)$ depends on the
transverse radial variable $r$ defined by
\be
r^3 = \sum_{i=1}^{4} \left\vert z_i \right\vert^2  \, .
\ee
The minimal value of $r$ is $r= \vert \epsilon \vert^{2/3}$ and in the deformed conifold case
it is useful to introduce a new coordinate $\tau$ defined by
\be
r^3 = \vert \epsilon \vert^2 \cosh (\tau)\ ,  \, \quad \tau\ge 0\ .
\ee
We refer the reader to the original paper \cite{Klebanov:2000hb} for the explicit
form of the deformed conifold metric, fluxes and the warp factor.

The field theory dual of the singular ($\epsilon=0$) conifold background
is a ${\cal N}=1$ $4d$ gauge theory at an IR conformal point, known as the Klebanov-Witten (KW)
theory \cite{Klebanov:1998hh}.
%The gauge group is $SU(N)\times SU(N)$.
%The bi-fundamental matter fields $A_{1,2}$, $B_{1,2}$ transform as $SU(2)\times SU(2)$ doublets and interact
%with a quartic superpotential
%\be  \label{Wkw}
%W_{\rm KW} =\lambda \epsilon^{i j}\epsilon^{k l} {\rm Tr}  \left( A_i B_k A _j B_l \right) \, .
%\ee
%The $U(1)$ isometry of the conifold corresponds to the R-symmetry. All bi-fundamentals have $U(1)$
%charge $1/2$ and the dimension of $3/4$.
%The gauge theory enjoys an additional $\mathbb{Z}_2$ symmetry that exchanges the gauge group factors and
%also the $A_i$'s with the $B_i$'s.
For the deformed conifold ($\epsilon \neq 0$)
%the conformal symmetry is broken and
the gauge group is
%now
$SU \left((k+1)M \right)\times SU(k M)$ for integer $k$.
The shift $k \to k-1$ describes a single step in a cascade of Seiberg dualities, which reduces the
gauge group down to $SU(2 M) \times SU(M)$ in the deep IR.
The R-symmetry is broken down to ${\mathbb Z}_2$ due to the
formation of a gluino condensate $\langle\lambda\lambda\rangle\sim\Lambda^3$.
On the gravity side the scale $\Lambda$ is fixed by the conifold
deformation parameter $\epsilon$.

At the next step we consider a $D7$-brane embedded in the KS background. The $D7$-brane extends along the
flat Minkowski space-time and a holomorphic non compact four cycle on the conifold $\Sigma$.
The embedding we are interested in is
\be  \label{z4mu}
z_4 = \mu \, .
\ee
It was shown in \cite{Kuperstein:2004hy} that the $D7$-brane wrapping $\Sigma$  with trivial world-volume
gauge field does not break the $\mathcal{N}=1$ SUSY of the original theory.
%Importantly, the brane configuration preserves the diagonal global $SU(2)_D$ which crucially simplifies the calculations.
Moreover, the $4$-cycle wrapped by the brane is topologically trivial \cite{Kuperstein:2004hy,Kuperstein:2008cq},
thus the tadpole cancellation condition is satisfied and no anti $D7$-brane is
required in order cancel the net RR charge. On the gauge theory side it implies that the
setup has no chiral symmetry. This is in contrast to the conifold based model of
\cite{Kuperstein:2008cq, Dymarsky:2009cm} (see also \cite{Bayona:2010bg,Ihl:2010zg}).

The lowest value of the radial coordinate $\tau_{\rm min}$ along the profile  depends
on the relative phase between $\mu$ and $\epsilon$.
A simple computation shows that
\be  \label{taumin}
\cosh \left( \tau_{\rm min} \right)  = \left\vert x^2 \right\vert +
                                   \left\vert 1 -  x^2 \right\vert \, ,
\ee
where
\be
x \equiv \frac{\mu}{\epsilon}\ .
\ee
This means that for real $x$ we have $\tau_{\rm min}=0$ as long as $|x| \leqslant 1$.
On the other hand, if $x$ has a non-zero imaginary part, then necessarily $\tau_{\rm min} > 0$.
For fixed $\tau$ the geometry of $\Sigma$ is $S^3/\mathbb{Z}_{2}$ which shrinks to $S^2$ at
$\tau=\tau_{\rm min}$.

\FIGURE[!b]{
\centering{
\includegraphics[]{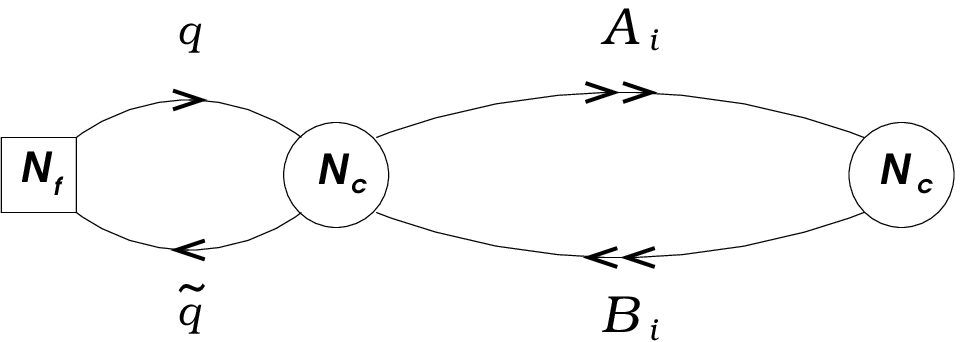}
\caption{The quiver diagram of the flavored KW model. For the KS model the ranks of the two gauge groups are different
and are not invariant under Seiberg duality. Yet, the structure of the quiver diagram
survives the cascade of Seiberg dualities.}
\label{Quiver}
}
}

The quiver diagram for the flavored Klebanov-Witten model is shown on Figure \ref{Quiver}
and the
%new
superpotential is \cite{Ouyang:2003df,Bigazzi:2008zt,Bigazzi:2008ie}
\be \label{W}
W = \lambda \epsilon^{i j}\epsilon^{k l} {\rm Tr}  \left( A_i B_k A _j B_l \right) - 2 \tilde{q} (A_1 B_1 + A_2 B_2 - \mu)  q - ( \tilde {q} q)^2 \, ,
\ee
where  $A_{1,2}$, $B_{1,2}$ are bi-fundamental matter fields.
Thus the term $ 2 \mu \tilde{q}q$ is responsible for the quark mass.

The right degrees of freedom for a confining theory are the di-baryon
operators $\mathcal{A}$ and $\mathcal{B}$ constructed from the bi-fundamentals or mesonic operators of the form
${\rm Tr} \left( A_i B_j \right)$. One of the striking features of the KS model
is that the superpotential retains its form under Seiberg duality.
The KS solution describes an $\mathcal{A} \leftrightarrow \mathcal{B}$-symmetric point in the baryonic branch,
which is evident, in particular, from the additional un-broken $\mathbb{Z}_2$ symmetry of the background.

Let us now address the map between various fields on the $D7$-brane and the gauge theory states.

As we have already mentioned in the Introduction, the world-volume fields on the flavor $D7$-brane are
a complex scalar $\delta\mu$ and a $8d$ gauge field $a_M$. Throughout this paper we will adopt the $a_\tau=0$ gauge.
This leaves the $4d$ $a_{\mu}$ components and the $a_{i}$ legs along $\Sigma$ at finite $\tau$.
The KS background describes a confining theory and only the colorless combinations
of the quarks fields, like the  mesons $q \tilde{q}$, appear in the spectrum.
Both $q$ and $\tilde{q}$ are $4d$  chiral multiplets. The bottom component of the operator $\tilde{q}q$
has dimension $3/2$. This is evident from the superpotential \eqref{W}.
This operator corresponds to the combination $a_1+i a_2$.
The top component $\int d^2\theta\ q \tilde{q}$ has dimension $5/2$ and corresponds to the complex scalar describing
 the embedding. Furthermore, the world-volume gauge field with space-time indexes $a_\mu$
is dual to the $U(1)_{\rm f}$ vector current $J_\mu$, and the remaining component of the world-volume
gauge field $a_3$ is dual to the bottom component of the $U(1)_{\rm f}$ current multiplet.

%So far we considered only the $SU(2)$-invariant modes that are associated with the radially-dependent perturbation of $\delta\mu$
%and $a_\mu,a_i$ on the $D7$-brane's world-volume.
%In general the $SU(2)$ charged modes have the same mass scale as the uncharged ones and should be taken into account.
%They would correspond to the operators of the form
%\bea
%\label{chargedmesons}
% q  A_{i_1} B_{i_2}... A_{i_{n-1}} B_{i_{n}} \tilde{q}
%\eea
%and such. Our main reason to omit these modes from the discussion is
% to make a connection with QCD, which has no adjoint or bi-fundamental matter and hence mesons of the form
%(\ref{chargedmesons}). Therefore in what follows we can forget about the angular coordinates and assume
%that all the fluctuations have only  dependence on the radial coordinate $\tau$.

In the holographic approach the $4d$ mesons arise as the normalizable fluctuations of
the $D7$-brane world-volume fields. The $\mathcal{N}=1$ SUSY arranges the particles into multiplets
combining a massive vector with a real scalar or
two real scalars together (we are not concerned with fermions here). There are no massless mesons in the setup in
question.  As was mentioned above the supersymmetry combines the vector mesons (the focus of this publication)
 with the scalars originating from the fluctuation of the world-volume gauge field $a_3$. Although these particles
have the same spectrum the equations they obey in the bulk are not the same.
Rather they are related to each other through a supersymmetric quantum
mechanics (SUSY QM) transformation \cite{Dymarsky:2010}.
This situation is typical for the fluctuations that are dual to operators of
different dimension. The equations for different particles in the multiplet coincide only if the corresponding
dual operators are of the same dimension. In the present case the vector current has dimension 3 and the bottom
component of the current multiplet has dimension 2, and therefore only the related spectra will coincide
but not the equations themselves. The SUSY QM transformation also relates the pair of scalars
originating  from the fluctuations of the world-volume gauge field $a_{1,2}$ and the fluctuations of the
embedding $\Re\ \delta \mu,\Im\ \delta \mu$. Since $a_{1,2}$ correspond to the operators of dimension $3/2<2$,
to find the spectrum one has to impose the ``unusual'' boundary conditions at
infinity \cite{Klebanov:1999tb}. While the general asymptotic of $a_{1,2}$ is $c_0r^{-1/2}+c_1 r^{-3/2}$,
the proper boundary conditions for $a_{1,2}$ are $c_1=0$ with arbitrary $c_0$.
In practice this would mean that to obtain the spectrum from the original equations for $a_{1,2}$ in the bulk,
using (for example) the shooting technique,   may not be the optimal strategy. Rather one can study the equations
for the superpartners $\Re\ \delta \mu,\Im\ \delta \mu$ which yield the same spectrum.
These fluctuations correspond to the operators of dimension $5/2>2$ and although they have the same  (up to an overall $r^{-1/2}$ factor) asymptotic at infinity $c_0+c_1 r^{-1}$,
they satisfy the conventional boundary conditions $c_0=0$ with arbitrary $c_1$.
It happens that in the conformal KW case there is an accidental degeneracy between
$a_1$ and $a_2$ (as well as between $\Re\ \delta \mu$ and $\Im\ \delta \mu$)
\cite{Dymarsky:2010}. It will be interesting to see if this degeneracy
will be lifted in the KS case when the new parameter $\epsilon$ is introduced.

\section{The spectrum calculation} \label{Numerics}

In this section we will calculate the spectrum of the vector mesons as a function of $\mu,\epsilon$.
These mesons are created by the $U(1)_{\rm f}$ flavor current $J_{\mu}$
dual to the Minkowskian vector field living on the $D7$-brane.
The five-dimensional action describing the lowest KK modes of this vector field has the form
\be\label{ac}
{\mathcal L} \sim C(\tau) F_{\mu \nu} F^{\mu \nu} + 2 D(\tau) F_{\mu \tau} F^{\mu}_{\tau} \, .
\ee
Here $\mu,\nu$ are the Minkowski indexes and $\tau$ is the radial direction along the conifold.
We choose the gauge $a_\tau=0$ and look for a solution of the form
$a_\mu = v_\mu e^{i k \cdot x} \psi(\tau)$ with  $v \cdot k=0$ and
\be  \label{spectrumEq}
	\left( D(\tau) \psi^\prime(\tau) \right)^\prime + m^2 C(\tau) \psi(\tau)=0 \, .
\ee
The four-dimensional meson mass $m^2 = -k^2$ emerges from the eigenvalue problem  (\ref{spectrumEq}).
In general, to  find the functions $C(\tau)$ and $D(\tau)$ could be a tiresome task.
This is because both the DBI and the CS parts of the $D7$-brane action contribute to the action (\ref{ac}).
The problem drastically simplifies for a SUSY $D7$-brane embedded into a constant dilaton, imaginary self-dual
(ISD) background \cite{Dymarsky:2009cm}.
In this case the invariant  2-form
\mbox{$\mathcal{F} \equiv \varphi^\star(B_2)+2\pi\alpha' d A$}, where $\varphi^\star$ denotes the pull-back,
satisfies  the {anti-self-duality} condition
\be
\label{ASD}
\mathcal{F} = - \star_4 \mathcal{F}\ ,
\ee
which implies \cite{Dymarsky:2009cm}
\be
C(\tau, \mu , \epsilon) = h \sqrt{g} - {\rm Pf} (\mathcal{F})\ , \qquad \textrm{and} \qquad
D(\tau, \mu , \epsilon) = g^{\tau \tau} \sqrt{g} \,.
\ee
Here $g$ stands for the un-warped induced metric on the 4-cycle $\Sigma$ wrapped by the $D7$-brane
and $\star_4$ is the Hodge dual associated with $g$.
The equation \eqref{ASD} follows directly
from the $\kappa$-symmetry condition, which in turn
ensures that the embedding preserves the background supersymmetry.
It was shown in \cite{Kuperstein:2004hy} that the $B$-field in the original gauge of the KS solution fulfills the $\kappa$-symmetry condition with zero world-volume gauge field $A=0$. Hence in what follows we can substitute
$\mathcal{F}$ with $B$ assuming the pullback on $\Sigma$.

As a result we find for our embedding
\begin{eqnarray}  \label{gB}
g^{\tau \tau} \sqrt{g} &=& |\epsilon|^{4/3} \frac{(\cosh(\tau) \sinh(\tau) - \tau)^{1/3} }{ 2\sinh^2 (\tau)}
                                  \left( \sinh^2(\tau) - 2 |x|^2 \cosh(\tau) + x^2 + \bar{x}^2 \right)\ ,     \\ \nonumber
\sqrt{g} &=& |\epsilon|^{8/3} \frac{(\cosh(\tau) \sinh(\tau) - \tau)^{2/3} }{ 8 \sinh^3(\tau)}
    \Bigg[| x \cosh(\tau) - \bar{x} |^2 +                                                          \\ \nonumber
    &&   \qquad \qquad  + \frac{ 2 \sinh^3(\tau) }{ 3(\cosh(\tau) \sinh(\tau) - \tau)}  \left(\sinh^2(\tau) - 2|x|^2\cosh(\tau) + x^2+\bar{x}^2 \right) \Bigg]\ , \\ \nonumber    {\rm Pf}B &=& -\frac{2^{1/3}}{8} \frac{|\epsilon|^{4/3}}{m_{\rm gb}^2}
       \frac{\left( \tau \cosh(\tau)-\sinh(\tau) \right)^2 }{ 2\sinh^5(\tau)} \left| x \cosh(\tau) - \bar{x} \right|^2\ , \,
\end{eqnarray}
and the KS warp factor is
\be
h(\tau) = \frac{|\epsilon|^{-4/3}}{m_{\textrm{gb}}^2}  I(\tau),
\qquad
\textrm{where}
\quad
I(\tau) = \int_\tau^\infty d z \frac{z \coth z -1}{\sinh^2 z} (\sinh(2z)-2z)^{1/3}  \, .
\ee
Here $m_{\rm gb}$ is the scale mass of the gauge theory glueballs
\be
m_{\textrm{gb}} = \frac{ \left\vert \epsilon \right\vert^{2/3}}{2^{1/3} g_s M \alpha^\prime} \, ,
\ee
and $x$ is the complex dimensionless parameter defined above
\be  \label{x}
x \equiv \frac{\mu}{\epsilon}\ .
\ee
To derive the first expression in (\ref{gB}) one has to introduce a complete set of local coordinates on $\Sigma$,
for example as it was done in \cite{Benini:2009ff}. The last two relations easily follow from the
explicit form of the K\"ahler form and the NS flux on the deformed conifold as well as the relations between different
forms on $\Sigma$ \cite{Dymarsky:2009fj}.

The mode $\psi$ from  \eqref{spectrumEq} is dual to an operator of dimension 3, a fact which is in agreement
with its asymptotic behavior at infinity\footnote{
Notice that at infinity $r^3 \approx \frac{1}{2}\left|\epsilon\right|^2 e^\tau $. }
\bea
\psi(\tau)  \approx c_0 +c_1 \cdot e^{-2\tau/3}\ .
\eea
The proper boundary conditions are $c_0=0$, \emph{i.e.} $\displaystyle{\lim_{\tau \to \infty} \psi(\tau) = 0}$
at infinity, and regularity at the origin
$\tau=\tau_{\rm min}$, where $\psi \approx 1+d_2 (\tau-\tau_{\rm min})^2$ with a numerical
coefficient $d_2=d_2(x, m/m_{\rm gb})$.
These boundary conditions represent no challenge for numerical studies of the spectrum
and the equation  \eqref{spectrumEq} can be solved with a standard shooting technique.
What we actually calculate in this way is the ratio $m/m_{\rm gb}$ between the meson and the glueball
masses as a function of $x$. Since the glueball mass $m_{\rm gb}$ is $\mu$ independent, changing
$x$ is equivalent to changing $\mu$, while keeping $\epsilon$ fixed.

\TABLE[!ht]{
\label{TableNum}
\centering{
\begin{tabular}{|r|cccccc|}
\hline
$x$ & \multicolumn{6}{|c|}{$m/m_{\rm gb}$}  \\
\hline
 & $n=1$ & $n=2$ & $n=3$ & $n=4$ & $n=5$ & $n=6$  \\
\hline
  0 & $\mathbf{1.487}$  & 2.774 & 4.077 & 5.385 & 6.696 & 8.009 \\
\hline
0.15& 1.488 & 2.773            & 4.073            & 5.379            & 6.688            & 7.999             \\
0.3 & 1.490 & 2.771            & 4.061            & 5.360            & 6.661            & 7.965             \\
0.45& 1.497 & 2.767            & 4.041            & 5.326            & 6.615            & 7.907             \\
0.6 & 1.504 & 2.763            & 4.014            & 5.277            & 6.547            & 7.820             \\
0.75& 1.516 & $\mathbf{2.763}$ & 3.981            & 5.211            & 6.451            & 7.697             \\
0.9 & 1.531 & 2.773            & $\mathbf{3.957}$ & $\mathbf{5.139}$ & $\mathbf{6.326}$ & $\mathbf{7.520}$  \\
1.2 & 1.575 & 2.880            & 4.134            & 5.395            & 6.665            & 7.941             \\
1.5 & 1.629 & 3.043            & 4.427            & 5.822            & 7.226            & 8.633             \\
1.8 & 1.687 & 3.208            & 4.701            & 6.203            & 7.708            & 9.216             \\
3   & 1.909 & 3.794            & 5.626            & 7.454            & 9.279            & 11.105            \\
7   & 2.506 & 5.249            & 7.858            & 10.450           & 13.014           & 15.584            \\
\hline
0.3 $\cdot$ $i$ & 1.497 & 2.802 & 4.120 & 5.443 & 6.767 & 8.094 \\
0.6 $\cdot$ $i$ & 1.525 & 2.878 & 4.236 & 5.599 & 6.964 & 8.328 \\
0.9 $\cdot$ $i$ & 1.566 & 2.985 & 4.401 & 5.820 & 7.239 & 8.659 \\
1.2 $\cdot$ $i$ & 1.614 & 3.108 & 4.590 & 6.073 & 7.557 & 9.041 \\
1.8 $\cdot$ $i$ & 1.718 & 3.373 & 4.994 & 6.613 & 8.234 & 9.851 \\
3   $\cdot$ $i$ & 1.928 & 3.886 & 5.777 & 7.660 & 9.541 & 11.421 \\
7   $\cdot$ $i$ & 2.512 & 5.279 & 7.905 & 10.503 & 13.093 & 15.679 \\
\hline
\end{tabular}
\caption{The vector meson spectrum for real and imaginary $x$. The lowest value of the mass
is indicated in bold.}
}
}

We solved \eqref{spectrumEq} numerically for various values of $x$. Because the spectrum is
invariant under $x\rightarrow -x$ and $x\rightarrow x^*$ we focused only on the $\Re\ x,\Im\ x>0$ quadrant.
In Table \ref{TableNum} we present our results{\footnote{
The mass values for $x=0$ are different from those in \cite{Kuperstein:2004hy} due to numerical errors therein.}
for $x=0$ and some real and purely imaginary values of $x$.
As one can see, if $x$ is real, for all modes, except for the lowest one,
the meson masses decrease as $x$ grows, until they reach their minimum value around $x \approx 1$.
Then they start growing again.
This type of behavior is not observed for the purely imaginary values of $x$, for which the masses
grow monotonically with $|x|$.
The results are presented graphically on Figures \ref{plot3d}, \ref{mureal}.
Our analysis seems then to imply that, for real $x$ smaller than $1$, the higher level mesons composed of
heavier quarks are actually lighter than those composed of lighter quarks.

\FIGURE[!ht]{
\centering
\includegraphics[width=0.42\textwidth]{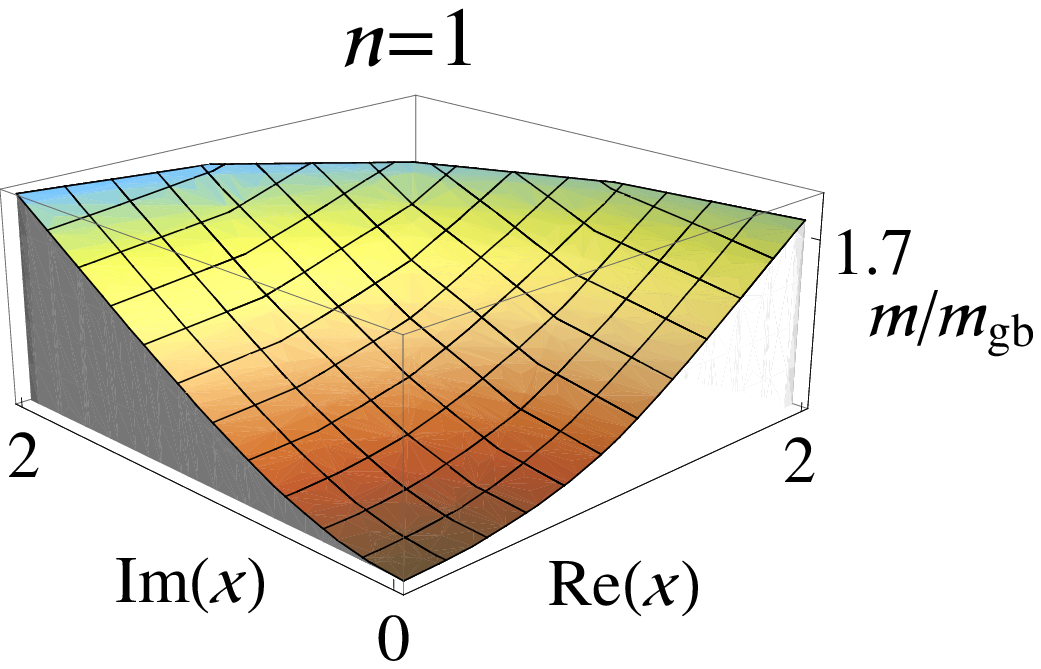}
\includegraphics[width=0.42\textwidth]{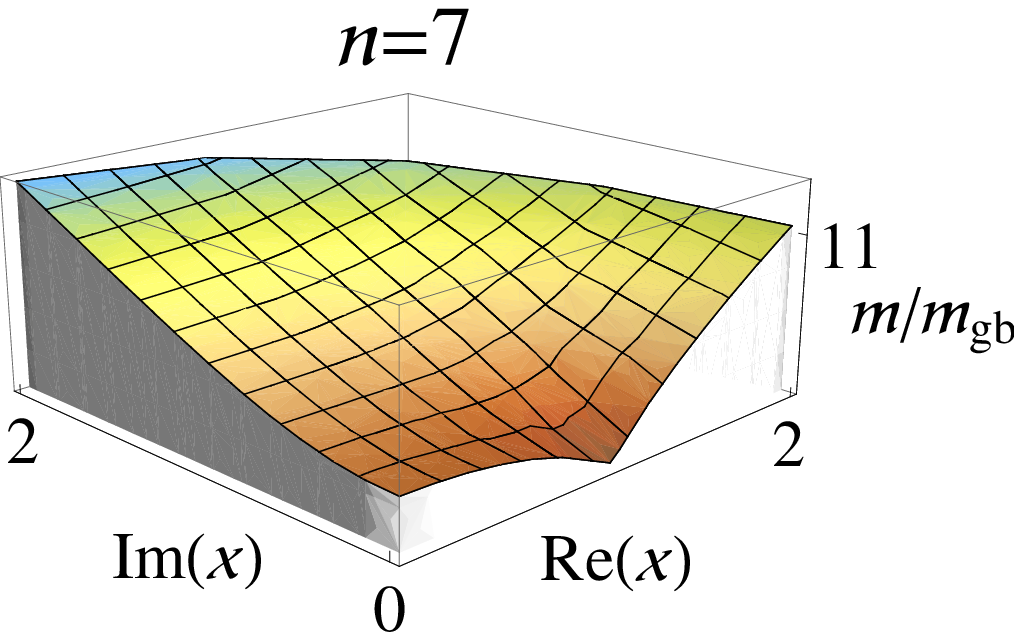}
\caption{Meson masses $m/m_{\rm gb}$ as functions of $x=\mu/\epsilon$ for $n=1$ (left) and $n=7$ (right).}
\label{plot3d}
}

This surprising behavior is confirmed by the WKB analysis, performed along the lines of \cite{Russo:1998by}
(as reported in Appendix B of \cite{Bigazzi:2009gu}).
After putting equation (\ref{spectrumEq}) in the  Schr\"odinger form
\begin{equation}
\partial_\tau^2 \phi(\tau) - V(\tau, x, m/m_{\rm gb})  \phi(\tau) = 0\ \, , \quad
V(\tau, x, m/m_{\rm gb}) =
  \frac{ \partial^2_\tau \left(  \sqrt{D(\tau, \mu , \epsilon)} \right)}{ \sqrt{D(\tau, \mu , \epsilon)} }
   - m^2 \frac{ C(\tau, \mu , \epsilon) }{ D(\tau, \mu , \epsilon) } \ ,
\end{equation}  one can use the standard WKB formula
\begin{equation}\label{generalWKB}
\left( n-\frac{1}{4} \right) \pi  = \int^{\tau^*}_{\tau_{\rm min}}d t \sqrt{-V(t, x, m/m_{\rm gb})}\, ,
     \qquad {\rm for } \quad n\geqslant 1\, .
\end{equation}
Here  $\tau^*$ is the point where the potential vanishes, $V(\tau^*)=0$. For larger $m$ this happens at
larger $\tau$ where $V(\tau)$ can be approximated by
\be
V(\tau, x, m/m_{\rm gb}) \approx \frac{1}{9} - \left( \frac{m}{m_{\rm gb}} \right)^2 \frac{(4\tau-1)}{16} e^{-2\tau/3} +
     \mathcal{O} ( e^{-\tau} ) \, .
\ee
Thus $\tau^*$ is very large for large $m$ and is $x$-independent. Hence we simply substitute it by infinity.
In the large $m$ limit we therefore have
\begin{equation}
\label{mass-n}
\frac{m_n}{m_{\rm gb}} \sim \frac{\pi}{\Delta(x)} n \, , \quad {\rm where} \quad
\Delta(x) = \int^{\infty}_{\tau_{\rm min}}  \sqrt{\Theta(\tau,x)}  d\tau  \quad {\rm and} \quad
\Theta(\tau,x) =  \frac{1}{m_{\rm gb}^2} \frac{C(\tau, \mu , \epsilon)}{D(\tau, \mu , \epsilon)} \, .
\end{equation}
In Figure \ref{mureal} we plot the meson masses calculated for real (left) and imaginary (right) $x$ for
the excited mode $n=7$; the bottom solid line is the WKB result, while the upper dashed line is  the result of
the shooting. The strange behavior found for real $x$ with the shooting technique is perfectly reproduced by
the WKB approximation.

A similar behavior was found in the model studied in \cite{Paredes:2006wb} (Figure 4).
It is worth emphasizing that also in that model the bare quark mass can be tuned to be smaller than the dynamical generated scale.
This is definitely not a common feature in holographyc models.
For example, in the original Sakai-Sugimoto model \cite{ss} it is unclear how to tune the bare mass (the distance of the flavor branes from the tip of the cigar just reflecting variations of the constituent mass).
In this sense, the KS background provides a privileged environment for the study of flavors with non-zero bare mass.

\FIGURE[!ht]{
\centering{
\includegraphics[width=0.45\textwidth]{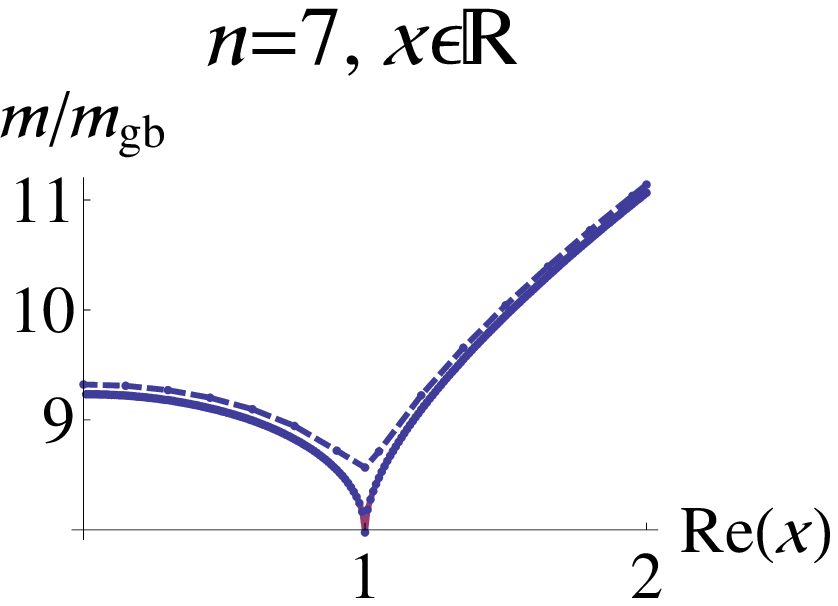}
\includegraphics[width=0.45\textwidth]{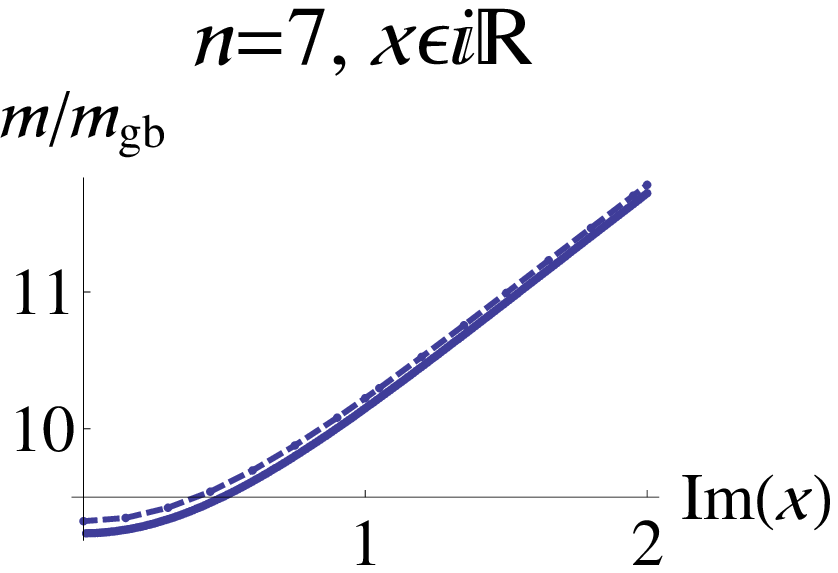}
\caption{Meson masses for real (left) and imaginary (right) $x$ for $n=7$ from the
WKB approximation (solid line) and the shooting technique calculation (dashed line).}
\label{mureal}
}
}

We can derive the strange behavior of the excited ($n>1$) modes for real
$x$ directly from the form of $\Theta(\tau)$ in (\ref{mass-n}). In the interval
$0\leqslant x < 1$ we have $\tau_{\rm min}=0$ and
the function $\Theta(\tau,x)$ has the same asymptotic values at the origin $\Theta(0,x)=2^{-5/3}3^{-1/3}I(0)$
and infinity $\displaystyle{\lim_{\tau \to \infty}} \Theta(\tau,x)=0$,
but it monotonically increases with $x$ for any $\tau$ in between zero and infinity
as demonstrated on the left graph on Figure \ref{thetaplot}.
Hence the integral $\Delta(x)$ is a monotonically decreasing function
with $x$, as expected. At $x=1$ the asymptotic value of $\Theta(\tau,x)$
at the origin $\tau=0$ jumps to $\Theta(0,1)=2^{-2/3}3^{-1/3}I(0)$
but $\Delta(x)$ is continuous. For $x>1$ the asymptotic behavior of $\Theta(\tau,x)$ changes drastically:
it blows at $\tau=\tau_{\rm min}$ as $\left( \tau-\tau_{\rm min} \right)^{-1}$.
It is clear though that for $x$ very close to $1$ the behavior of $\Theta(\tau,x)$ as a function of $\tau$
will be the same as for $x=1$ except for some very small vicinity of $\tau=\tau_{\rm min}$.
Hence the integral $\Delta(x)$ is a continuous function of $x$
as is expected from the physics of the problem. On the right graph on Figure \ref{thetaplot} we adjusted the minimal value of the coordinate $\tau$
to the origin, $ \tilde{\tau} \equiv \tau - \tau_{\rm min}$.
It is clear now that as $x$ increases the function $\Theta(\tilde{\tau},x)$
decreases at each $\tilde{\tau}$. Therefore $\Delta(x)$
is monotonically decreasing with $x$ when $x\sim 1$.
As a result we find that $\Delta(x)$ reaches its maximum,
and the masses of the excited modes reach their minimum, at $x=1$.
This is close to the numerical results in Table \ref{TableNum}, where the lowest masses of the
excited modes appear around $x=0.9$. It is worth mentioning here that the discussion
in this paragraph is based on the WKB approximation which is usually not valid for small $n$, especially for $n=1$, and
so the monotonic behavior of the lowest mode cannot be addressed in this framework.

\FIGURE[!ht]{
\centering{
\includegraphics[width=0.48\textwidth]{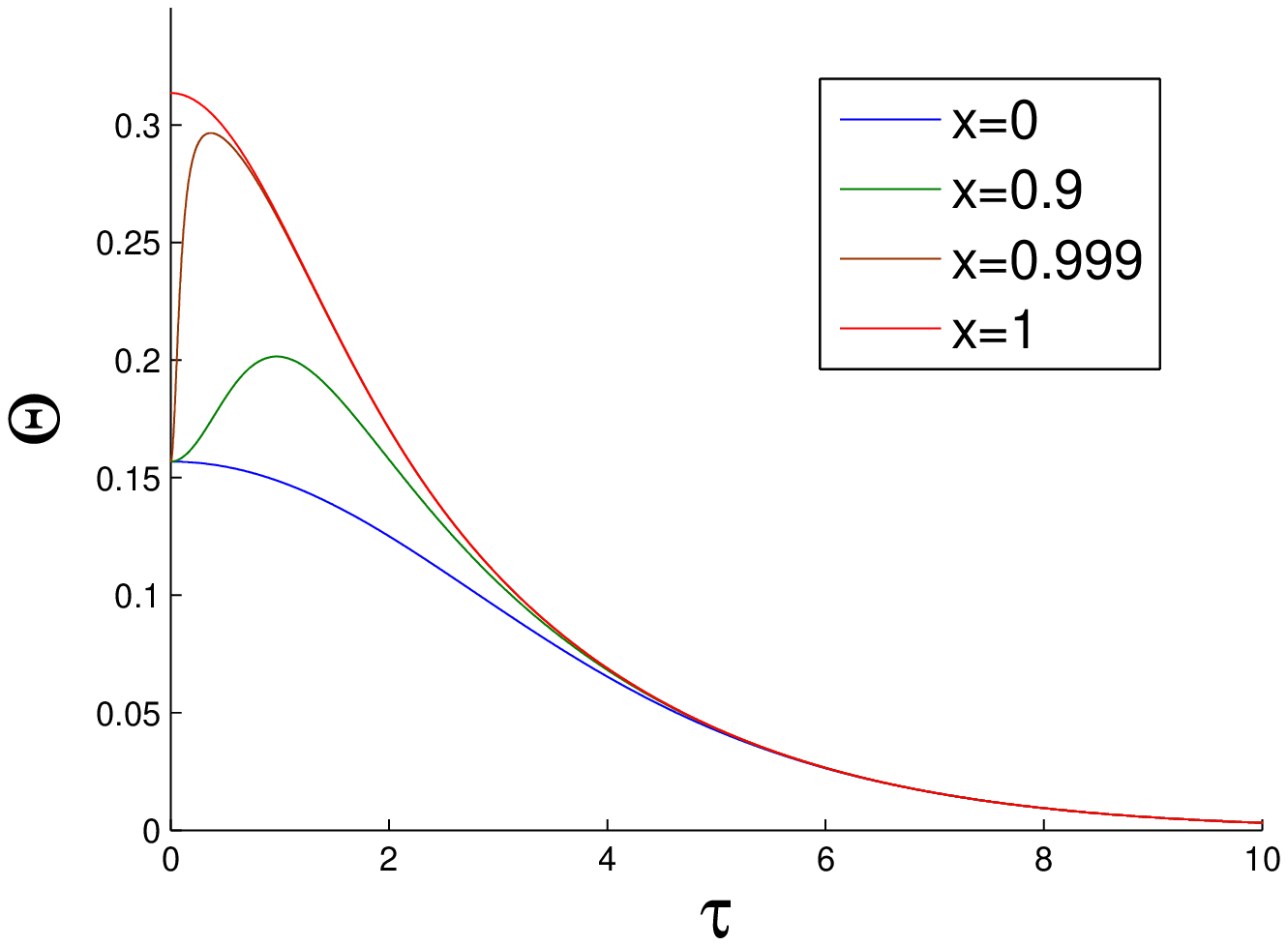}
\includegraphics[width=0.48\textwidth]{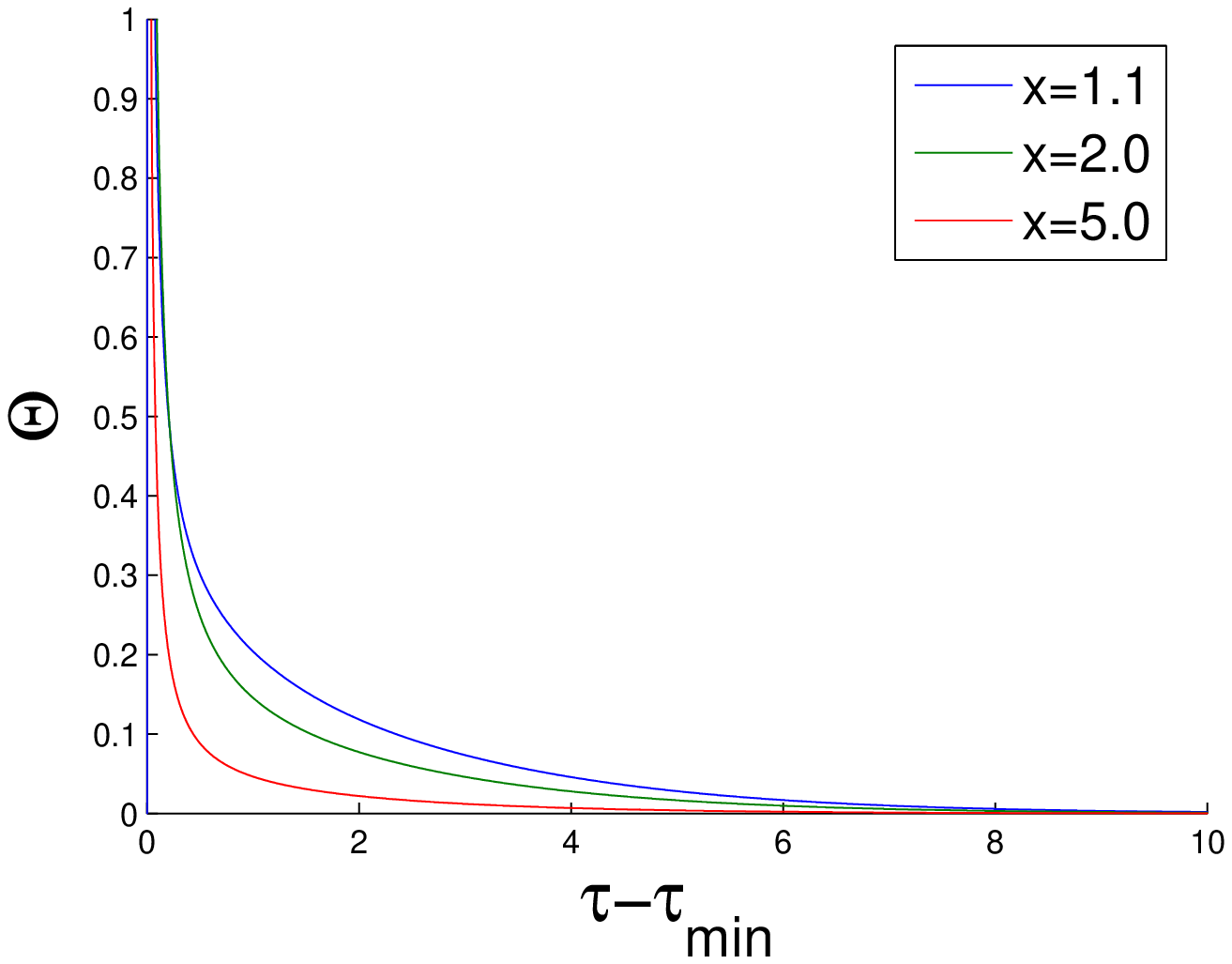}
\caption{The function $\Theta(\tau,x)$ for $x \leqslant 1$ (left) and $x>1$ (right).}
\label{thetaplot}
}
}

The WKB analysis  can also help us to see how the $\epsilon$ dependence disappears in the KW limit when $x\gg 1$.
Indeed in this case we expect the meson mass to depend only on $\mu$ and hence to be proportional to
$m_{\rm gb} |x|^{2/3}$.
Such a behavior readily follows from \eqref{mass-n}. $\Theta(\tau,x)$ goes as $e^{-2 \tau/3}$ for large $\tau$
and in order to find $\Delta(x)$ we need to integrate the square root of $\Theta(\tau,x)$
from $\tau_{\rm min}$ to infinity. For large $|x|$  the parameter $\tau_{\rm min}$ is also large and
can be approximated by $e^{\tau_{\rm min}} \approx 4 |x|^2$ as follows from \eqref{taumin}.
Putting all together we find $\Delta(x)$ to be proportional to $|x|^{-2/3}$
and therefore according to \eqref{mass-n} we get $m_n \approx m_{\rm gb}|x|^{2/3} n$ as expected.

\subsection*{A possible explanation}  \label{Explanation}

Before proceeding to the comparison with the known experimental data in the next section we would like to discuss
the possibility that the strange behavior of the meson spectrum in our setup is a result of some intrinsic features
of our model that are not present in QCD.

Our proposal here is to attribute the observed phenomenon to the quartic interaction between the quarks and the bi-fundamental fields in the superpotential \eqref{W}. Intuitively this interaction may modify the ``effective" quark mass
and hence the meson masses. Since we work in the quenched approximation, we might ignore the dynamics
of the bi-fundamentals. Being more precise, in order to estimate the ``effective" quark mass, we substitute the VEVs of $A_i, B_i$ into \eqref{W}.
The VEVs are estimated as follows.

The moduli space of the gauge theory is in one to one correspondence with the geometry of the
deformed conifold. This fact is helpful in finding the superpotential \eqref{W} for the conformal KW theory. But the
interpretation of \eqref{W} for the cascading KS theory is more tricky since the bi-fundamentals
$A_i$ and $B_i$ are not the good degrees of freedom anymore. In the original paper \cite{Klebanov:2000hb}
the ``mesonic" fields $M_{i j} = \textrm{Tr} (A_i B_j)$ were proposed as the low-energy fields in the dual gauge theory.
It was later realized in \cite{Aharony:2000pp} (see \cite{Dymarsky:2005xt} for the detailed discussion)
that the theory is actually on its baryonic branch, $M_{ij}=0$ and the low-energy fields are the di-baryon operators constructed from the bi-fundamental fields ${\mathcal A},{\mathcal B}\sim |A|^{M},|B|^M$.
Although this step is not well-defined, we can evaluate the VEV of the original bi-fundamental fields
$A_i$ and $B_i$ to be roughly of order $\sqrt{\epsilon}$. This follows from the simple dimensional analysis.

Thus, in order to estimate the ``effective" quark mass, we should plug $\epsilon$
instead of  $A_i B_j$ in  \eqref{W}.
The substitution gives roughly $ \tilde{q} (A_1 B_1 + A_2 B_2 - \mu)  q \rightarrow  \tilde{q} (\epsilon - \mu)  q$, so that
%Then
the effective quark mass is
\be
\label{meff}
m_{\rm eff} \sim \left\vert \epsilon - \mu \right\vert =
                                   \left\vert \epsilon \right\vert \cdot \left\vert 1 - x \right\vert \, .
\ee
As we increase $\mu$ keeping $\epsilon$ fixed, $m_{\rm eff}$ strongly depends on the relative phase between $\mu$
and $\epsilon$. In particular, for real $x=\mu/\epsilon$ the mass $m_{\rm eff}$ decreases as we increase $x$ all the way
until we reach $x=1$.
This  is closely related to the geometry of the $D7$-brane profile for real and imaginary $\mu$ as
can be seen from the similarity  between \eqref{meff} and \eqref{taumin}.
Now, assuming that the usual dependence of meson mass on $m_{\rm eff}$ (i.e. it increases when $m_{\rm eff}$ increases) we recover  the unusual behavior of the excited modes for real $0<x<1$.
An analogous explanation was suggested in \cite{Paredes:2006wb}.

In the light of this argument
it is actually the behavior of the  lowest mode in Table \ref{TableNum} that needs an explanation.
Say, the lowest mode was precisely massless for $\mu=0$, \emph{i.e.} it was a Goldstone boson of a certain spontaneously
broken symmetry. Then for $\mu\neq 0$ the symmetry would be broken explicitly and the Goldstone boson would get
a mass proportional to $\mu$. Such a situation is realized in the model  of \cite{Kuperstein:2008cq} where the
two massless modes correspond to the broken chiral and conformal symmetries. In our case there is no spontaneously
broken symmetry at $\mu=0$. The behavior of the lowest mode, though, reminds of a \emph{pseudo} Goldstone boson.
This is similar to the situation with the pions in QCD,
which are the pseudo Goldstone bosons of the chiral symmetry explicitly broken by the quark masses.

\section{Comparison to the experimental data}  \label{Experiment}

In this section we compare our results with the spectrum of the $\rho$ and $\phi$ mesons, \emph{i.e.} the light quark $1^{--}$ vector mesons in QCD.\footnote{The heavy quark $1^{--}$ mesons are not tightly bound in QCD,
so the comparison with our values would be pointless.}
Before proceeding let us recall that the correspondence
\be  \label{correspondence}
\epsilon^{2/3} \sim \Lambda \, , \qquad\qquad \mu^{2/3} \sim m_{\rm q}\, ,
\ee
is not rigorous. We just assume that the constant of proportionality is of the same order in the both cases.
Similarly to $\epsilon$, the power of $\mu$ here follows merely from dimensional analysis.

Throughout this section we will use the standard $n^{2s+1}l_j$ notations for the $q \bar{q}$
mesons \cite{Eidelman:2004wy}. Here $n$ stands for the radial excitation number,
$s$ is either $0$ (anti-parallel quark spins) or $1$ (parallel quark spins), and $l$
is the relative angular momentum of the quarks inside the meson.
The latter is usually denoted by the conventional  notations $S$, $P$, $D$, $\ldots$ for $l=0,1,2,\ldots$
Finally,  $\vert l-s \vert \leqslant j \leqslant l+s$ is the total angular momentum of the state.
An additional notation for the mesons is $j^{PC}$, where $P$ and $C$ are the parity and the charge
conjugation respectively. In terms of $l$ and $s$ one has $P=(-1)^{l+1}$ and $C=(-1)^{l+s}$,
and so the $S$-wave ($l=0$) $\rho$-mesons, that we study in this paper, have $1^{--}$ quantum numbers.

\subsection*{$1\%$ errors for the $\rho$-meson masses}

The masses and widths of the first $\rho$ mesons in units of MeV are \cite{Eidelman:2004wy}
\begin{eqnarray}\label{rhos}
\rho(770): &  775.49 \pm 0.34 \quad & (\Gamma=149.1)\, , \nonumber \\
\rho(1450):&  1465 \pm 25    \quad & (\Gamma=400)\, , \\
\rho(2150):&  2149 \pm 17    \quad & (\Gamma=359)\, .\nonumber
 \end{eqnarray}
The first two are reported as $1\, ^3 S_1$ and $2\, ^3 S_1$ modes; the latter is not firmly identified, but we report it for the reasons that will become apparent in a moment.
We did not include the $\rho(1700)$ because it is the $1\, ^3 D_1$ mode, and as such is not supposed to be captured by our analysis (the modes with higher spin in holographic models are parametrically heavier and  correspond
to stringy states; they are not captured by the supergravity approximation). We also ignored all heavier
modes whose status is unclear.

We consider the two ratios
\begin{equation}\label{ratiosrho}
\frac{m_{\rho(1450)}}{m_{\rho(770)}}\sim 1.88(9)\ , \qquad \frac{m_{\rho(2150)}}{m_{\rho(770)}}\sim 2.77(2)\ .
\end{equation}
The $\rho$-mesons are mixtures of the $u$ and $d$ quarks.
Hence the quark mass scale is $m_u\sim m_d \sim (m_u+m_d)/2\sim 3.8 \, {\rm MeV}$.
For the dynamical scale we can use $\Lambda\sim m_{\rho(1450)}\sim 775 \, {\rm MeV}$
as usual in the $\pi$-meson physics.\footnote{$\Lambda$ is supposed to set the scale for the meson resonances in QCD, modulo small corrections due to the finite quark masses, so a fair estimate is given by the first vector meson mass.}
We will assume that $3.8/775$ is the counterpart of the geometrical parameter $x^{2/3}$.
As we mentioned above the relation is not precise.
Nevertheless, the basic point is that the resulting geometric ratio $x$ is clearly very small, to the extent that using
different prescriptions for  $x$ would not produce a sizable difference.\footnote{For example, using
$\Lambda\sim 200 \, {\rm MeV}$ we get  $x\sim 0.0026$ but the resulting difference in the meson mass
ratios is only of order $10^{-3}$.}
We are therefore allowed to use the $x=0$ results from Table \ref{TableNum} for the comparison
\begin{equation}\label{ratiosx0}
\frac{m_{n=2}}{m_{n=1}}\sim 1.86(6)\ , \qquad \frac{m_{n=3}}{m_{n=2}}\sim 2.74(2)\ .
\end{equation}
The deviation of the values in (\ref{ratiosx0}) from the experimental data in (\ref{ratiosrho}) are about $1\%$ in both cases. In particular it is of the same order as the experimental error. It is now apparent why we included the $\rho(2150)$ meson: the agreement of its mass with the mass of our third mode is so precise that we are led to believe that $\rho(2150)$ is really the $3\, ^3 S_1$ mode.

It is important to emphasize here that the status of the
$3\, ^3 S_1$ mode is currently unclear (see for example \cite{Klempt:2007cp}).
Although an additional resonance(s) (with the same quantum numbers) between $\rho(1450)$ and $\rho(2150)$
has by now been observed by many groups,
its nature and properties, however, are far from being clear.
This mode, sometimes called $\rho(1900)$, with a predicted mass of about $1830$ MeV, should exist if the spectrum of the light $m_{\rm q} \ll \Lambda$ mesons (like the ones we study) is consistent with the \emph{linear} confinement, namely if $m_n^2 \sim n$.
If indeed the linear relation $m_n^2 \sim n$ is correct then the $\rho(2150)$ meson
will be the $4\, ^3 S_1$ mode. This would reduce the precision of our results.

The idea that $m_n^2$ grows as $n$ can be supported by a simple semiclassical argument (see \cite{Karch:2006pv}
for the basic review and the related AdS/CFT discussion). There are also some experimental indications
that the masses of the light mesons satisfy the linear $m^2_n\sim n$ relations \cite{Klempt:2007cp, Anisovich:2000kxa}.
In the holographic models, however,
$m_n^2$ usually grows like $n^2$ \cite{Kruczenski:2003be}.
In our case this is evident from the WKB approximation \eqref{mass-n}.
The possible exception being the lowest
modes for which $m^2_n\sim n^2$ does not necessarily hold. Hence the holographic models are not expected to describe the highly excited states with large $n$, but theoretically could give a reasonable prediction for the lowest states.

Leaving aside the $\rho(1900)$ issue, the surprising matching of the holographic calculation
with the experimental meson masses is {a priori} unjustified.
A list of reasons why our analysis may not be applicable includes the fact that the Klebanov-Strassler theory is
supersymmetric, it contains many more light fields than QCD, has a large number of colors, and so on.\footnote{It is probably not even in the same universality class \cite{Gubser:2004qj}.
The recent paper \cite{Domokos:2011dn} discusses 
another source of discrepancy between the two theories, \emph{i.e.} the effect 
on the $\rho$ spectrum of the mixing of the vector and tensor quark 
bilinears in QCD, which is not captured by standard holographic models.
}
Given these fundamental differences,  one possibility is that the remarkable agreement observed above
happens by chance. Another explanation would be that, once the masses are normalized by the lightest one,
these ratios are in fact robust and are not sensitive to the differences between the holographic model and
QCD. Clearly, this can be true only for the few lightest modes before the different  $n$ dependence kicks in.
Let us note here that a similar matching for the ratios of masses of the light(est) glueball states in the same
holographic model and the pure glue $SU(3)$ theory on the lattice was observed in \cite{Dymarsky:2008wd}.

\subsection*{Higher meson mass inversion}\label{inversionqcd}

The comparison with the $\phi$ mesons is more complicated, since the holographic calculation would correspond to  the ``pure" $s \bar s$ states, while the real $\phi$ mesons can be mixtures of the $s$ with the $u, d$ quarks.
We will use a prescription from \cite{Allton:1996yv,Iatrakis:2010zf,Iatrakis:2010jb} to give an
estimate for the mass of a putative ``pure" $s \bar s$ mesons (which we will denote as ``$\phi$"-mesons
in what follows\footnote{As in \cite{Iatrakis:2010zf} the quotation marks are to remember that these are not the real particles.}). The mass of  ``$\phi$" is given by twice the mass of the corresponding $K^*$ meson
(the light-strange vector meson) minus the mass of the $\rho$ meson (the light-light)
meson.\footnote{It is an implicit assumption here that the binding energy is  more or less the same for both mesons.}

The masses of the first two $K^*$ mesons are
\begin{eqnarray}
K^*(892):&  896 \pm 0.25\quad & (\Gamma=50.3)\, , \nonumber \\
K^*(1410):&  1414 \pm 15\quad & (\Gamma=232)\, ,
 \end{eqnarray}
and we obtain for the ``$\phi$"'s
 \begin{eqnarray}\label{phis}
``\phi(1020)":&&  1015 \pm 0.5\, , \nonumber \\
``\phi(1680)":&&  1363 \pm 33\, .
 \end{eqnarray}
Note that the mass of the real $\phi(1020)$ is very close to the putative $``\phi(1020)"$ one, consistently with
the expectation of a very small mixing of the $s \bar s$ state with the $u \bar u$, $d \bar d$ for the ground state
meson.
Instead, the mass of the excited $\phi(1680)$ is quite larger than the one of the putative $``\phi(1680)"$,
a fact which agrees with a possible sizable mixing in this case.

>From these considerations we immediately obtain one of the main results of this note: the lowest mode of ``$\phi$''
in (\ref{phis}) is heavier
than that one of $\rho$ in (\ref{rhos}) as expected, but the first excited mode of ``$\phi$"  is \emph{lighter}
than the one of the $\rho$-meson.
This reproduces the behavior observed in Section \ref{Numerics} for real $x\leqslant 1$.
Indeed, one can see from Table 1 that the second (first excited) mode ($n=2$) of our $\rho$ meson, (\emph{i.e.}
essentially the $x=0$ entry equal to $2.774$) is heavier than the second mode of our ``$\phi$" mode (some entry
for small $x$, e.g. for $x=0.15$ the entry is equal to $2.773$).

A quick look at the measurement errors shows that they are not sufficient to account for this behavior in QCD.
The difference in masses is not small,  it is of order of $5-10\%$ of the meson masses,
and it is hard to attribute it to a mere coincidence.
Of course, we are just considering putative ``pure" $s\bar s$  mesons and, most importantly, the widths
of such states are typically very large. Nevertheless, the effect described above is also quite large,
so we believe that the qualitative behavior of our holographic model makes sense and it is actually not
excluded that it could even be found in Nature.

As an aside, let us add that if we want to compare the QCD ``$\phi$" masses with the ones of the holographic
model at hand we need to know the precise mapping of the geometric parameter $\mu/\epsilon$ to the field theory
value of $m_{\rm q}/\Lambda$.
With all the reasonable choices, the discrepancy between our results and the QCD ones is larger than the one found
for the $\rho$ mesons and of the typical order for holographic computations applied toward phenomenology (5\%-50\%).
Yet the crucial phenomena stressed above -- the mass inversion of the higher meson modes \emph{i.e.}  that the
higher mesons made of heavy quarks tend to be lighter than those made of lighter quarks, is not sensitive to the
details of this map.

Let us conclude this section by mentioning that in \cite{Bigazzi:2009gu, Bigazzi:2008qq} the screening effects of
light dynamical flavors on the spectrum of mesons composed of massless quarks in the Klebanov-Strassler theory has
been analyzed.
Along the same lines, it would be interesting to see  the effects of light dynamical flavors on the results above
and in particular on the observed behavior of the meson masses for real $x\leqslant 1$.

\section*{Acknowledgements}

We are grateful to Francesco Bigazzi for collaboration on the early stages of this project.
It is a pleasure to thank E. Klempt, J. Maldacena, D. Melnikov, A. Paredes and J. Soto for useful discussions,
and J. Sonnenschein for his comments on the final version of the manuscript.
We are grateful to E. Kiritsis for correspondence regarding the $\rho$-meson spectrum.

A.D. thanks the theory group at Vrije University Brussel for hospitality while this work was initiated.
The research of A.D. was supported by the Stanford Institute for Theoretical Physics, by the DOE grant
DE-FG02-90ER40542, by the Monell Foundation, and in part by the grant RFBR 07-02-00878 and the Grant for Support of
Scientific Schools NSh- 3035.2008.2.

The work of S.~K. is supported by the European Commission Marie Curie Fellowship under the
contract IEF-2008-237488.

The research leading to the results in this paper has received funding from the European Community's Seventh Framework Programme (FP7/2007-2013 under grant agreement n. 253534).
This work is also supported by the FWO -Vlaanderen, project G.0235.05 and by the Federal Office for Scientific, Technical and Cultural Affairs through the Interuniversity Attraction Poles Programme (Belgian Science Policy) P6/11-P.

{ \it A. L. C. would like to thank the Italian students, parents, teachers and scientists for
their activity in support of public education and research.}

%\section*{\bf Appendix}

\bibliographystyle{utphys}
\bibliography{VectorMesonsSpectrum}

\end{document}